\documentclass{ifacconf}

\usepackage{graphicx}      % include this line if your document contains figures
\usepackage{natbib}        % required for bibliography
\usepackage{mathrsfs}
\usepackage{amssymb}
\usepackage{color}
\usepackage{mathtools}
\usepackage{comment}
\usepackage{subfig}
\usepackage{float}
\usepackage{tabularx}
\usepackage{makecell}
\usepackage{caption}
\usepackage{url}
\usepackage[short]{optidef}
\usepackage[linesnumbered,ruled,vlined,algo2e]{algorithm2e}
\usepackage{algorithmic}
\usepackage{mathptmx}
\usepackage[11pt]{moresize}
\usepackage{bbold}
\usepackage{amsmath}

\newtheorem{definition}{\textbf{Definition}}

\newcommand{\ignore}[1]{}

\begin{document}
\begin{frontmatter}

\title{An Iterative Method to Learn a Linear Control Barrier Function} 

% \thanks[footnoteinfo]{Sponsor and financial support acknowledgment
% goes here. Paper titles should be written in uppercase and lowercase
% letters, not all uppercase.}

\author[First]{Zihao Liang} 
\author[First]{Jason King Ching Lo} 
% \author[First]{Shaoshuai Mou}

\address[First]{School of Aeronautics and Astronautics,
        Purdue University, IN 47907, USA (e-mail: liang331@purdue.edu, jasonlkc724@gmail.com).}

% \begin{abstract}                % Abstract of not more than 250 words.
% Control barrier function is a common and effective approach for enforcing and validating the safety requirements for safety-critical systems. However, constructing such function for a general system is a non-trivial task. This paper proposes an iterative, optimization-based framework to obtain control barrier functions (CBFs) for various applications of safety controller synthesis. We consider a known user-specify safety measures for autonomous systems which does not necessarily render its own zero-superlevel set forward invariant; this superlevel set can be a known set of allowable states for an autonomous system that is obvious given the context of the task. The CBF is parameterized as a set of linear functions of states. By taking samples from the set of allowable states described by the given safety measures, an optimization problem is formulated to solve for the linear function coefficients such that they render the safe set forward invariant. In addition, our framework explicitly addresses control input constraints during the construction of CBFs. We demonstrate the effectiveness of our framework by synthesizing the learned CBF into an existing controller of the nonlinear Moore-Greitzer jet engine model in no-stall model and analyze the results. The effects of different levels of input constraints are also analyzed and the corresponding visualizations of the CBFs are presented. 
% \end{abstract}

\begin{abstract}                % Abstract of not more than 250 words.
Control barrier function (CBF) has recently started to serve as a basis to develop approaches for enforcing safety requirements in control systems. However, constructing such function for a general system is a non-trivial task. This paper proposes an iterative, optimization-based framework to obtain a CBF from a given user-specified set for a general control affine system. Without losing generality, we parameterize the CBF as a set of linear functions of states. By taking samples from the given user-specified set, we reformulate the problem of learning a CBF into an optimization problem that solves for linear function coefficients. The resulting linear functions construct the CBF and yield a safe set which has forward invariance property. In addition, the proposed framework explicitly addresses control input constraints during the construction of CBFs. Effectiveness of the proposed method is demonstrated by learning a CBF for an nonlinear Moore Greitzer jet engine, where the system trajectory is prevented from entering unsafe set.
\end{abstract}

\begin{keyword}
Control Barrier Functions, Nonlinear Systems, Data-driven Control.
\end{keyword}

\end{frontmatter}
%===============================================================================

\section{Introduction}
Safety guarantee is one of the most crucial factor that needs to be considered before deploying an autonomous system into the real world. It is necessary to integrate safety criteria into controller design process. For example, safety restraint has to be applied to a self-driving car in order to kept the human unharmed. Moreover, damage to the system and the environment should be avoided. Designing such controller for a safety-critical autonomous system is nontrivial. 

Recently, control barrier functions (CBF) become a popular choice to enforce safety on autonomous systems.  Barrier functions were originally developed to address the safety requirements of a system (\cite{Nagumo1942berDL,SafetyVerificationHybridSys,BarrierCertificates}). This idea was extended to consider controls and the notion of "control barrier function" was formally defined in (\cite{FirstCBFDef}). Recently the work in (\cite{CBFcruiseControl,CBFbasedQP}) redefined the definition of \emph{control barrier functions (CBF)} by extending Nagumo's theorem to the entire safe set (rather than just on the boundary). An important consequence is that one can now construct safe controller using such CBF since it is now defined at all points within the set. This is typically done through a \emph{min-norm} controller that alters the nominal control in a minimally invasive manner (\cite{CBFbasedQP}). In (\cite{CBFBipedalrobot,3D_dynamicWalkingSteppingStone,PreciseFootStep}), CBFs are applied to develop safe controller for bipedal robots and enable them to walk on stepping stones. In automotive applications, CBF has be used to provide guarantees on safety features such as lane-keeping and adaptive cruise control (\cite{CBFcruiseControl,CorrectnessGuaranteesLaneKeeping,mobileRobotTestbeds}). Safety-critical controllers for aerial system such as quadrotors have also been synthesize using CBFs (\cite{planarQuadrotor,DifferentialFlatnessQuadrotors}). Experimental work has been done to prove the effectiveness of using CBF to control non-statistically stable systems such as a Segway in (\cite{FrameworkforSafetyCriticalActiveSetInvariance}). The application of CBF also extends into multi-robot systems (\cite{SwarmBehavior,CollisionFreeMultiRobot,ConnectivityMaintenance,RobotEcology}), where safe maneuvers and methods for motion coordination are developed.   

It is evident that control barrier functions has a wide range of applications and its ability to provide provable safety guarantee is desirable. However, it is not clear how one can find such CBF and its corresponding safe set for an arbitrary control system. This can, in general, be a difficult problem since CBF is very application-dependent.

Without knowing the algebraic form of the CBF, one cannot leverage the safety guarantee that provided by the CBF. There are several recent works investigated this problem via learning. In (\cite{SafeLearningQuadrotor}), a CBF is incrementally learned to cope with unmodeled disturbance, with a conservative CBF provided at the beginning of the learning process. Authors of (\cite{NeuralCertificates}) introduce a neural-network-based approach to jointly learn the control policy along with CBF and Lyapunov function. This approach requires prior knowledge of safe and unsafe sets. In (\cite{SVM}), with given safe and unsafe data, support vector machine is used to obtain a CBF. For (\cite{NeuralCertificates}) and (\cite{SVM}), the proposed methods rely on given classified data from safe set and unsafe set, which could be difficult to collect. Authors of (\cite{NeuralCertificates,ExpertDemonstrations,ROBEY20211}) propose an optimization framework which has a cost function that includes barrier functions on CBF conditions. In (\cite{ConstrainedMotionPlanning}), the CBFs are parameterized by linear functions. The proposed approach incrementally updates the linear functions from human demonstrations. These aforementioned methods generally neglect the control constraints, and therefore the property of forward invariance can almost always be rendered. In reality, this can be a problem since there are actuator constraints and input bounds that limit the system's capability. For this reason, one must make an effort to ensure the input constraints do not conflict with the required condition on the time derivative of CBF. The authors of (\cite{FixedWingAircraft}) consider the input constraint. The authors assume a given feedback control law, which is somewhat limited. Moreover, the CBF generated from a given allowable set only guarantees safety in finite time.  

In general, CBF is obtained based on safety requirements or user-specified performance. For example, in order to guarantee two fixed-wing aircraft will not collide, (\cite{FixedWingAircraft}) considers a performance measure that is equal to the squared distance between the two aircraft in excess of some minimum safety distance. However, this performance measure itself does not render a set forward invariant, therefore is not a CBF. In this paper, we assume given data from an allowable set defined by some performance measure. In applications, it could be the point cloud data obtained using Lidar on the autonomous vehicles. Such performance measure is tailored for different applications. Due to the input constraints considered, not all points in the allowable set can always be rendered safe and therefore one cannot say the predefined performance measure is a control barrier function. Specifically, the introduction of input constraints may cause certain points in the allowable set to lose property of forward invariance, e.g., the input that is required to keep the system in the safe may lie outside of the input constraint.

This motivates us to find a function such that all the states within its zero-superlevel set can be rendered forward invariance with the given input constraint and a performance measure. Once such function is found, we can claim such function to be a valid control barrier function and the safe set is indeed the zero-superlevel set of this function.

Without losing generality, we employ a linear approximation rather than using neural networks to approximate a nonlinear CBF (\cite{NeuralCertificates,ExpertDemonstrations,ROBEY20211}). Using neural networks requires a set of data that is already labelled safe or unsafe. Compared to linear functions, neural networks have stronger expressive power and need longer computation time to converge. Since this paper set out to propose an iterative, optimization-based method to compute CBFs, computation time is a crucial factor and is prioritized when considering different representation methods. Moreover, many CBF applications typically involve nonlinear functions, which means nonlinear programming problems need to be solved. Parameterizing the CBF with linear functions is an attempt to convert the nonlinear problem into a linear programming problem and hopefully decreases the computational cost.

The main contribution of this paper is an iterative, optimization-based framework for obtaining control barrier functions with control input constraints from a given set of allowable states defined by application specific performance measure. The control barrier function is parameterized using a collection of linear equations and the corresponding safe invariant set is represented as the intersection of half spaces defined by the linear equations. We also present validation for the learned control barrier function as well as a series of simulations to show the effectiveness of our method.

\section{Problem Formulations}

\begin{comment}
As seen in the aforementioned work, safety can often be expressed through the notion of set invariance. In particular, by restricting the system to remain within a chosen safe set one can guarantee safety of a system. Consider a safe set $\mathcal{C} \subset \mathcal{D}$ defined to be the zero-superlevel set of a continuous differentiable function $h: \mathcal{D} \subset \mathbb{R}^n \Rightarrow \mathbb{R}$, e.g.,
\begin{gather}
    \mathcal{C}\;=\;\{x \in \mathbb{R}^n | \;h(x) \geq 0\}, \\
    \partial \mathcal{C}\;=\;\{x \in \mathbb{R}^n | \; h(x) = 0\}, \\
    Int(C)\;=\;\{x \in \mathbb{R}^n | \; h(x) > 0\}.
\end{gather}
\end{comment}

% \subsection{System Dynamics}
Let us consider a nonlinear control affine system of the form
\begin{equation}
    \label{dynamics}
    \Dot{\boldsymbol{x}} = \boldsymbol{f}(\boldsymbol{x}) + \boldsymbol{g}(\boldsymbol{x})\boldsymbol{u},
\end{equation}
where $\boldsymbol{x}\in\mathcal{X}\subseteq\mathbb{R}^n$ denotes the system state, with $\mathcal{X}$ as the system state space. To account for the actuation limitations in a mechanical system, the control input $\boldsymbol{u}$ is constrained by $\mathcal{U}=\{\boldsymbol{u}| u_{\min}\leq \boldsymbol{u}\leq u_{\max}\}$, where $\leq$ denotes entry-wise inequality. 
Let $\boldsymbol f:\mathbb{R}^n \rightarrow \mathbb{R}^n$ and $\boldsymbol g:\mathbb{R}^{n} \rightarrow \mathbb{R}^{n\times m}$ be locally Lipschitz functions, and denote  $I(\boldsymbol{x}_0) = [0,\tau_{max})$ as the maximum interval of existence on which $\boldsymbol{x}(t)$ is a unique solution to (\ref{dynamics}) from the initial condtion $\boldsymbol{x}(0)=\boldsymbol{x}_0\in\mathcal{X}$.

As safety guarantee is one of the most crucial factors for a system, the notion of \emph{safety} for a system must be defined. Consider a continuously differentiable function $h:\mathcal{X}\subset\mathbb{R}^n \rightarrow \mathbb{R}$, and define a set $\mathcal{C}$ as the \emph{zero-superlevel} set of $h$:
\begin{equation}
    \mathcal{C}=\{x\in\mathcal{X}\subset\mathbb{R}^n:h(x)\ge 0\}\label{h>=0}.
    % \partial\mathcal{C}=\{x\in\mathcal{X}\subset\mathbb{R}^n:h(x) = 0\}\label{h=0}\\
    % Int(\mathcal{C})=\{x\in\mathcal{X}\subset\mathbb{R}^n:h(x) < 0\}\label{h<0}
\end{equation}

The set is considered as a \emph{safe set} if it satisfies certain user-specified safety requirements and can be rendered forward invariant:

\begin{definition} [Safety as forward invariant set]
    A set $\mathcal{C}$ is considered forward-invariant if for any initial state $\boldsymbol{x}_0 \in \mathcal{C}$, $\boldsymbol{x}(t) \in \mathcal{C}$, $\forall t \in I(x_0)$. If the state of a system starts in a safe set $\mathcal{C}$ always remains within $\mathcal{C}$, then the system is considered \emph{safe}. 
    \label{def1}
    (\cite{theory&app})
 \end{definition}

With \emph{safety} properly defined, the definition of control barrier function which will be used to enforce such property is presented: 

\begin{definition}[Control barrier function]
     Consider the control affine system (\ref{dynamics}) in a state space $\mathcal{X}$. A continuously differentiable function $h:\mathcal{X}\subset\mathbb{R}^n \rightarrow \mathbb{R}$ is a \emph{control barrier function (CBF)} if there exists an extended class $\mathcal{K}_\infty$ function \footnote{An extended class $\mathcal{K}_\infty$ function is a strictly increasing function $\alpha:\mathbb{R}\rightarrow\mathbb{R}$ with $\alpha(0)=0$.}  $\alpha$ such that 
    \begin{equation}
        \sup_{\boldsymbol{u} \in \mathcal{U}}[ L_f h(\boldsymbol{x}) + L_g h(\boldsymbol{x})\boldsymbol{u} \ge -\alpha(h(\boldsymbol{x}))], \quad \forall \boldsymbol{x} \in \mathcal{X}. \label{h_dotcondition}
    \end{equation}
    \label{def2}
    (\cite{theory&app}) 
\end{definition} 
Note $L_f h(\boldsymbol{x})$ and $L_g h(\boldsymbol{x})$ are Lie-derivatives of $h(\boldsymbol{x})$, e.g., $L_f h(\boldsymbol{x}) = \frac{\partial h(\boldsymbol{x})}{\partial \boldsymbol{x}} f(\boldsymbol{x})$ and $L_g h(\boldsymbol{x}) = \frac{\partial h(\boldsymbol{x})}{\partial \boldsymbol{x}} g(\boldsymbol{x})$. We also assume that $\frac{\partial h(x)}{\partial x}\neq 0$. 

\begin{comment}
In general, finding such a control barrier function and the corresponding safe-invariant set for an arbitrary system can be a difficult problem since it is application dependent. More importantly, current work that involve control barrier functions generally neglects the control constraints, e.g., $\mathcal{U}=\mathbb{R}^m$, and therefore the condition (\ref{h_dotcondition}) can almost always be satisfied. In reality, this can be a problem since there are actuator constraints and input bounds that limit the system. For this reason, one must make an effort to ensure the input constraints do not conflict with the (\ref{h_dotcondition}); in other words, the set of safe controllers $K_{cbf}(x) = \{u\in\mathcal{U}: L_f h(\boldsymbol{x}) + L_g h(\boldsymbol{x})u + \alpha(h(\boldsymbol{x}))  \geq 0\} \,\forall \boldsymbol{x} \in \mathcal{X}\}$ is not empty.Furthermore, the \emph{safe} set $\mathcal{C}$ is asymptotically stable in $\mathcal{X}$. \cite{theory&app} 
\end{comment}

Given a set $\mathcal{C}\subset\mathbb{R}^n$ defined in (\ref{h>=0}) by a continuously differentiable function $h:\mathcal{X}\subset \mathbb{R}^n\rightarrow\mathbb{R}$. If $h$ is a control barrier function on $\mathcal{X}$ then any Lipschitz continuous controller $\boldsymbol{u}\in\mathcal{U}$ fulfilling (\ref{h_dotcondition}) renders the set $\mathcal{C}$ \emph{safe}. 
% \label{CBFEnsuresSafety}

In this paper, we consider a known set of allowable states $\mathcal{X}_{\rho}$. Such set is obtained via safety requirements specified by the user and does not necessarily have the property of forward invariance. 

Suppose the control barrier function $\boldsymbol{h}(\boldsymbol{x})$ is parameterized using a collection of linear equations
\begin{equation}
\label{CBF of linear inequalities}
    \boldsymbol{h}_{\boldsymbol{A},\boldsymbol{b}}(\boldsymbol{x})=\boldsymbol{A}\boldsymbol{x}+\boldsymbol{b},
\end{equation}
where $\boldsymbol{x}\in\mathcal{X}_{\rho}$, $\boldsymbol{A}\in\mathbb{R}^{L\times n}$, $\boldsymbol{b}\in\mathbb{R}^{L}$, and $L$ is the number of linear equations. The safe set defined by the CBF is:
\begin{equation}
    \mathcal{X}_s=\{x\in\mathcal{X}_{\rho}:\boldsymbol{h}_{\boldsymbol{A},\boldsymbol{b}}(\boldsymbol{x})\ge 0\}\label{Xs}.
\end{equation}
where $\geq$ denotes entry-wise inequality.

% Our \textbf{problem of interest} is to obtain the parameter $\boldsymbol{A},\boldsymbol{B}$ of the paramterized control barrier function $\boldsymbol{h}_{\boldsymbol{A},\boldsymbol{B}}(\boldsymbol{x})$ from a known set of allowable states $\mathcal{X}_{\rho}$ with the consideration of control input constraints. The relationship between the CBF $h(\boldsymbol{x})$, safe set $\mathcal{X}_s$ and the given set of allowable states $\mathcal{X}_{\rho}$ is demonstrated in Fig. (\ref{fig:desiredCBF}).

\begin{figure}
    \centering
    \includegraphics[width=0.25\textwidth]{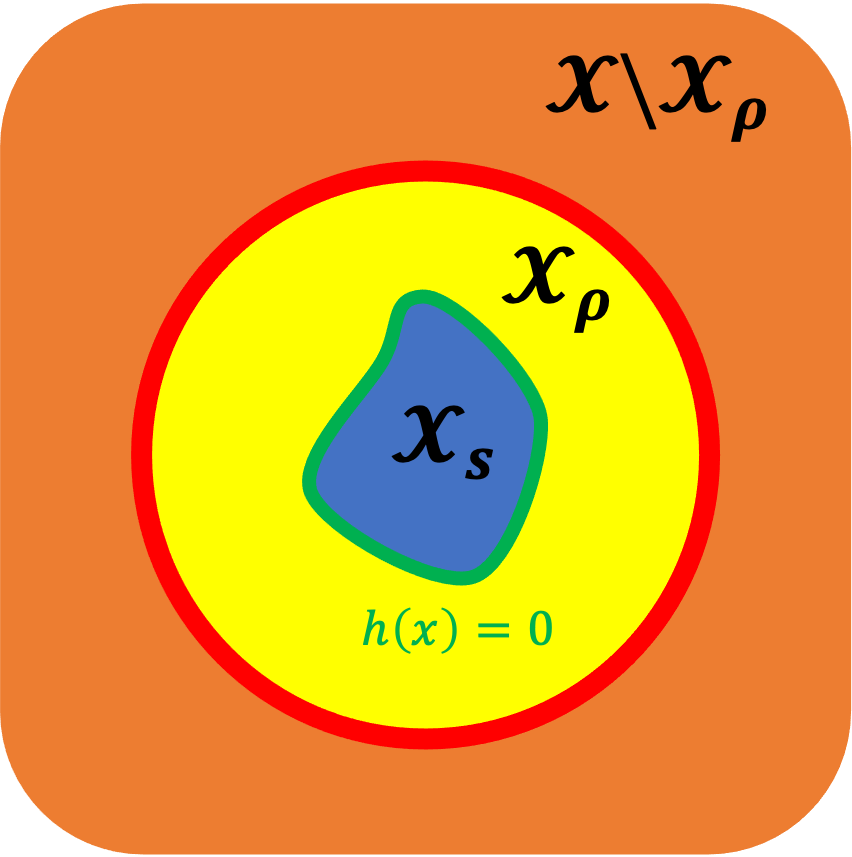}
    \caption{Illustration of CBF and corresponding safe-invariant set}
    \label{fig:desiredCBF}
\end{figure}

Our \textbf{problem of interest} is to obtain the parameter $\boldsymbol{A},\boldsymbol{b}$ of the paramterized control barrier function $\boldsymbol{h}_{\boldsymbol{A},\boldsymbol{b}}(\boldsymbol{x})$, such that its zero-superlevel set can be rendered forward invariant and will be a subset of $\mathcal{X}_{\rho}$, with the consideration of control input constraints. This superlevel set is then the safe set $\mathcal{X}_s$. The relationship between the CBF $h(\boldsymbol{x})$, safe set $\mathcal{X}_s$ and the given set of allowable states $\mathcal{X}_{\rho}$ is demonstrated in Fig. (\ref{fig:desiredCBF}).

\section{An iterative method to obtain CBF}

\subsection{The Sampling-based Optimization Problem}

To make this problem tractable and to efficiently search for a valid control barrier function, we parameterize such function using polytopic approximation similar to (\cite{ConstrainedMotionPlanning}). In particular, a collection of linear inequalities will be used to denote the control barrier function and resulting safe set is represented as the intersection of positive half spaces. Positive half spaces are defined by linear inequalities $h_i(\boldsymbol{x}) = \boldsymbol{a}_i^T\boldsymbol{x} + \beta_i\ge 0$, with $\boldsymbol a_i\in\mathbb{R}^n$, $\beta_i\in\mathbb{R}$; $i=\{1,\cdots,L\}$, where $L$ is the number of linear inequalities that are used to construct the safe set. The CBF defined by the collection of linear inequalities can then be expressed by
\begin{equation}
\label{CBF of linear inequalities}
    \boldsymbol{h}_{\boldsymbol{A},\boldsymbol{B}}(\boldsymbol{x})=\boldsymbol{A}\boldsymbol{x}+\boldsymbol{b},
\end{equation}
where 
\begin{align}
\label{AandB}
    \boldsymbol{A}= 
      \begin{bmatrix}
     \boldsymbol{a}_1^T\\
      \boldsymbol{a}_2^T\\
      \vdots \\
      \boldsymbol{a}_L^T\\
    \end{bmatrix}
    \in\mathbb{R}^{L\times n}, \qquad
    \boldsymbol{b}=
    \begin{bmatrix}
    \beta_1 \\ \beta_2 \\ \vdots \\ \beta_L\\
    \end{bmatrix}
    \in\mathbb{R}^{L}. 
\end{align}
Here, the CBF should possess the same properties as mentioned in (\ref{h>=0}-\ref{h_dotcondition}) according to (\cite{ConstrainedMotionPlanning}), although we note that the dimensions are different, where $\boldsymbol{h}_{\boldsymbol{A},\boldsymbol{b}}(\boldsymbol{x})\in\mathbb{R}^L$. 

% The safe set defined by (\ref{CBF of linear inequalities}) is as follow:
% \begin{equation}
%     \mathcal{X}_{s}=\{x\in\mathbb{R}^n:\boldsymbol{A}\boldsymbol{x}+\boldsymbol{b} \ge 0\},
%     \label{Ax+B>=0}
% \end{equation}
% where $\geq$ denotes entry-wise inequality.

\subsubsection{\textbf{Ensuring forward invariance with bounded inputs:}}
To ensure the learned function is a valid CBF, it must satisfy properties stated in (\ref{h_dotcondition}). The condition (\ref{h_dotcondition}) requires $\dot{\boldsymbol{h}}_{\boldsymbol{A},\boldsymbol{b}}(\boldsymbol{x}) \geq -\alpha(h(\boldsymbol{x})), \forall \boldsymbol{x}\in\mathcal{X}$, meaning there is always a feasible input that allows us to drive the system towards the interior of zero-superlevel of $\boldsymbol{h}_{\boldsymbol{A},\boldsymbol{b}}(\boldsymbol{x})$. In our case, we relax such condition and only require that $\dot{\boldsymbol{h}}_{\boldsymbol{A},\boldsymbol{b}}(\boldsymbol{x}) \geq -\alpha(h_{\boldsymbol{A},\boldsymbol{b}}(\boldsymbol{x})), \forall \boldsymbol{x}\in\mathcal{X}_\rho$. To ensure this forward invariant property, with the CBF represented by a group of linear inequalities, (\ref{h_dotcondition}) can be written in the form:
\begin{equation}
    \boldsymbol{A}\boldsymbol{f(x)}+\boldsymbol{Ag(x)u}\ge -\alpha(\boldsymbol{Ax+b}).
    \label{h_dot_linear}
\end{equation}

In addition, we also consider bounded control inputs $\mathcal{U}=\{\boldsymbol{u}| u_{\min}\leq \boldsymbol{u}\leq\ u_{\max}\}$, where $\leq$ denotes entry-wise inequality. The set of admissible inputs is represented by:
\begin{equation}
    \boldsymbol{C}\boldsymbol{u} \le \boldsymbol{D},
    \label{input_bound}
\end{equation}

where
\begin{equation}
    \boldsymbol{C}=
    \begin{bmatrix}
        I_m\\
        -I_m
    \end{bmatrix}
    \in\mathbb{R}^{2m\times m}, \quad
    \boldsymbol{D}=
    \begin{bmatrix}
    u_{max}\boldsymbol{1}_m^T \\-u_{min}\boldsymbol{1}_m^T\\
    \end{bmatrix}
    \in\mathbb{R}^{2m}.
\end{equation}

% \begin{equation}
%     \boldsymbol{C}=
%     \begin{bmatrix}
%         I_m\\
%         -I_m
%     \end{bmatrix}
%     \in\mathbb{R}^{2m\times m}, \quad
%     \boldsymbol{D}=
%     \begin{bmatrix}
%     u_{max}\\ \vdots \\ u_{max}\\-u_{min}\\\vdots\\-u_{min}\\
%     \end{bmatrix}
%     \in\mathbb{R}^{2m}.
% \end{equation}

% Moreover, the obtained CBF should also satisfy the following equation:
% \begin{equation}
%     \boldsymbol{A}\boldsymbol{x}+\boldsymbol{B} \ge 0,
%     \label{Ax+B>0}
% \end{equation}
% where $\geq$ denotes entry-wise inequality.

By stacking (\ref{h_dot_linear}) and (\ref{input_bound}), we can get:
\begin{equation}
    \bar{\boldsymbol{A}} \boldsymbol{u} \le \bar{\boldsymbol{B}}
    \label{forward_invariant_ineq}
\end{equation}

where
\begin{equation}
    \bar{\boldsymbol{A}}
    =
    \begin{bmatrix}
        \boldsymbol{-Ag(x)}\\
        \boldsymbol{C}
    \end{bmatrix},
    \bar{\boldsymbol{B}}
    =
    \begin{bmatrix}
         \boldsymbol{A}\boldsymbol{f(x)}+\alpha(\boldsymbol{Ax+b})\\
         \boldsymbol{D}
    \end{bmatrix}.
\end{equation}
%This needs to be worked on too
\begin{comment}
Claim: If it is asymptotically stable in $\mathcal{X}_\rho$, it is asymptotically safe w.r.t. $\mathcal{X}_u$ (guarantee won't enter $\mathcal{X}_u$). Rho sort of acts like a buffer region.
Make a simple proof to say if it cannot exit $\mathcal{X}_\rho$ then it won't ever enter $\mathcal{X}_u$.
\end{comment}
% We note that this not only ensures the invariance property of $\mathcal{X}_{s}$ but also makes it asymptotically stable in $\mathcal{X}_\rho$. 
We now sample the interior of the set of allowable states and denote these interior samples as $\boldsymbol{x}_k\in\mathcal{X}_{\rho}, k=\{1,...,M\}$.  For each sample point, the conditions (\ref{forward_invariant_ineq}) must be satisfied.
This constraint is introduced in coherence with (\ref{h_dotcondition}) to ensure the forward invariance of the obtained safe set. Therefore, $\bar{\boldsymbol{A}}$ and $\bar{\boldsymbol{B}}$ are functions of $\boldsymbol{x}_k$ and $\boldsymbol{u}_k$, i.e. $\bar{\boldsymbol{A}}(\boldsymbol{x}_k)$, $\bar{\boldsymbol{B}}(\boldsymbol{x}_k,\boldsymbol{u}_k)$, where $\boldsymbol{u}_k$ is the admissible input for $\boldsymbol{x}_k$ to enforce the condition (\ref{h_dot_linear}).  In conclusion, constraint (\ref{forward_invariant_ineq}) ensures the existence of an \emph{admissible input} value that can render the safe set $\mathcal{X}_{s}$ forward invariant.

\subsubsection{\textbf{Objective function:}}
The objective of this optimization problem is designed to avoid the overestimation of the volume of the safe set $\mathcal{X}_s$ defined by the obtained CBF $\boldsymbol{h}_{\boldsymbol{A},\boldsymbol{b}}$. The idea is to avoid having the safe set include states that do not belong to the given allowable set $\mathcal{X}_{\rho}$, which could lead to the violation of CBF condition.

As the safe set is defined by the inequality $\boldsymbol{A}\boldsymbol{x}_k+\boldsymbol{b}\ge 0$, the most conservative estimation of the CBF given the sample state $\boldsymbol{x}_k$ would be when $\boldsymbol{A}\boldsymbol{x}_k+\boldsymbol{b}= 0$. Therefore, to avoid overestimating the volume of the safe set $\mathcal{X}_{s}$, the objective function for the proposed optimization problem is designed as:
\begin{equation}\label{eqn:cost}
    J=\sum_{k=1}^{M} \|{\boldsymbol{A}\boldsymbol{x}_k+\boldsymbol{b}}\|_1.
\end{equation}
The proposed objective function indirectly represents the sum of distances between each sample state $\boldsymbol{x}_k$, and each linear boundary of the resulting safe set. Minimizing (\ref{eqn:cost}) yields a compact safe set that encloses the maximum number of sample states.

\subsubsection{\textbf{Optimization problem to obtain CBF:}}
%Talk about norm constraint
By integrating the aforementioned constraints and cost function, we now present the problem of learning a CBF as a sampling-based optimization problem. Let $M$ be the number of sample points generated from the interior of $\mathcal{X}_{\rho}$. The optimization problem is then set up as follows:

\begin{mini}
    {\boldsymbol{A},\boldsymbol{b},\boldsymbol{U}}{\sum_{k=1}^{M} \|{\boldsymbol{A}\boldsymbol{x}_k+\boldsymbol{b}}\|_1}{\label{Opt_Prob}}{}
    % \addConstraint{\min_{\boldsymbol{u}}(\boldsymbol{a}_i^T\boldsymbol{f}(\boldsymbol{x}_k)+\boldsymbol{a}_i^T\boldsymbol{g}(\boldsymbol{x}_k)\boldsymbol{u}+\alpha(\boldsymbol{a}_i^T\boldsymbol{x}_k+b_i))\le0}\label{invariantconstraint}
    \addConstraint{\bar{\boldsymbol{A}} \boldsymbol{u_k} \le \bar{\boldsymbol{B}},\quad\forall\boldsymbol{x}_k}\
    \addConstraint{\boldsymbol{A}\boldsymbol{x}_k + \boldsymbol{b}\ge0,\quad\forall\boldsymbol{x}_k,}
\end{mini}
where
\begin{equation}
    \boldsymbol{U}=[\boldsymbol{u_1},\boldsymbol{u_2},\cdots,\boldsymbol{u_k}]
\end{equation}

\subsection{An iterative method to obtain CBF}

To enforce safety on autonomous system, CBF is typically applied via the QP-CBF framework that is proposed in (\cite{CBFbasedQP}). In this framework, current state is given as a starting point of a trajectory in the safe set defined by CBF. According to the definition of safety with CBF, this trajectory must never leave the safe set. To more consistently obtain a feasible solution, we propose an iterative method for computing the CBF. We first represent the given allowable set as an undirected graph, where the vertices are equally spaced points in the allowable sets. Then, starting from the given current state, breadth-first-search (BFS) is used to traverse the allowable set, and the list of visited states is stored. The optimization problem ($\ref{Opt_Prob}$) is solved every time an allowable state (node) is visited and added to this list of visited states. The goal of the method is to obtain a CBF such that the volume of its safe set approaches the volume of the given allowable set. We use the number of allowable states included in $\mathcal{X}_{s}$ as an indirect measure of the safe set's volume. As more allowable states are visited, the feasible solution of the optimization problem ($\ref{Opt_Prob}$) should include more allowable states, and the volume of the safe set defined by the obtained CBF will increase and approach the volume of the given allowable set. After all allowable states are visited, the feasible solution of ($\ref{Opt_Prob}$) that defines a safe set with the largest volume is selected as the final solution of this method.

\subsection{Implementation Detail}
\subsubsection{\textbf{Initial guess of $\boldsymbol{A}$ and $\boldsymbol{b}$}:}
The safe set is initialized using a convex polytope that encloses all the visited states. To find such polytope, we utilize the Graham scan (GS) method, which gives a set of vertices that represent the polytope (\cite{Graham1972AnEA}). The set of vertices is then converted to a set of inequality constraints, i.e. $\boldsymbol{A}\boldsymbol{x}\ge\boldsymbol{b}$ ,which represents the polytope edges that most tightly enclose the visited points. These inequality constraints will then be used as the initial guess of the CBF.

The pseudo code of the proposed method is demonstrated in Algorithm \ref{algorithm}.

\IncMargin{1em}
\begin{algorithm2e}[h]
	\caption{Iterative Method to Obtain CBF} \label{algorithm}
	\DontPrintSemicolon
	\SetKwInOut{Input}{Input}
	\SetKwInOut{Output}{Output}
	\SetKwInput{Initialize}{Initialize}

	\Input{$\mathcal{X}_{\rho},\boldsymbol{x}_t$ {\ssmall //$\ $allowable set, current state.}}
	\Output{$\boldsymbol{A}_{opt},\boldsymbol{b}_{opt}$}
	\Initialize{$V=[\boldsymbol{x}_t]$, $F=[\boldsymbol{x}_t]$, $Q=[\boldsymbol{x}_t]$\; {\ssmall //$\ $list of visited states, list of feasible states, BFS queue}}
    $G=genGraph(\mathcal{X}_{\rho})$ {\ssmall //$\ $generate undirected graph for allowable set.}\;
    \While {$Q$ is not empty}{
    $v = Q.pop(1)$\;
    $N$ = neighbors($G$, $v$) {\ssmall //$\ $find neighbor states.}\;
    \For{$i=1:size(N)$}{
    \If{$N(i)$ is not in $visited$}{
        $Q.append(N(i))$\;
        $V.append(N(i))$\;
        {\ssmall //$\ $generate initial polytope using GS.}\;
        $[A, b]=genPoly(V)$\;
         {\ssmall //$\ $generate CBF by solving (\ref{Opt_Prob}).}\;
        $[A, b,flag]=genCBF(V,A,b)$\;
        {\ssmall //$\ $$flag=1$ indicates feasible solution.}\;
		\eIf{$flag == 1$}{
			$F=V$\;
            $A_{opt},b_{opt}=A,b$\;}
		{
		do nothing
		}
	}
	}
	}
\end{algorithm2e}
\DecMargin{1em}

\section{Numerical Results}
\subsection{Learning Control Barrier Function from Allowable Set}
\begin{figure}
    \centering
    \includegraphics[width=0.4\textwidth]{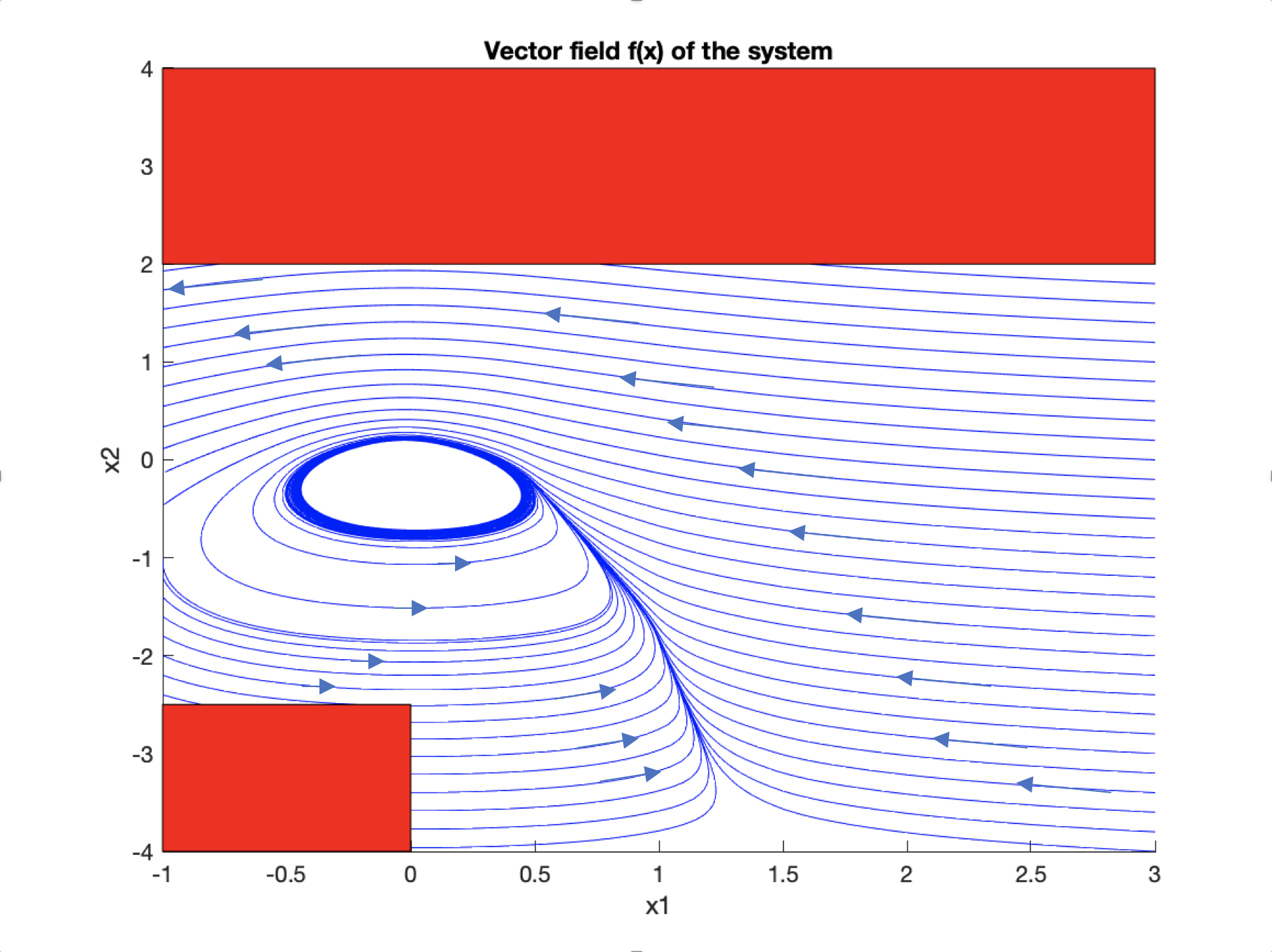}
    \caption{Vector field $f(x)$ of the system: $X_u$ are represented  by the red filled regions and the $X_\rho$ is the unfilled region}
    \label{fig:vectorField}
\end{figure}

To demonstrate the effectiveness of our method discussed in section III, we apply it to a real system and show how the obtained CBF can be used to keep the system from entering undesired states. We consider the nonlinear Moore-Greitzer jet engine model in no-stall mode used in (\cite{UnknownNonlinearGP}) as our example. The dynamics of this system is given as
\begin{equation*}
    \boldsymbol{f}(\boldsymbol{x}) = 
    \begin{bmatrix}
    f_1(\boldsymbol{x})\\
    f_2(\boldsymbol{x})
    \end{bmatrix} = 
    \begin{bmatrix}
    x_2 - \frac{3}{2} x_1^2 - \frac{1}{2} x_1^3 \\
    x_1
    \end{bmatrix}
    \text{and} \:
    \boldsymbol{g}(\boldsymbol{x}) = 
    \begin{bmatrix}
    0 \\
    -1
    \end{bmatrix}
    ,
\end{equation*}

where $\boldsymbol{x} = [x_1, x_2]^T$ represents our system states. Physically, $x_1 = 1 - \Phi$ and $x_2 = \Psi - \psi - 2$ are quantities proportional to the mass flow $\Phi$ and the pressure rise $\Psi$, respectively. We consider the state-space of interest $X = [-1,3] \times [-4,4]$ and the unsafe region $\mathcal{X}_u = [-1,0]\times[-4,2.5]\cup[-1,3]\times[2,4]$. The allowable set is then $\mathcal{X}_\rho = \mathcal{X} \setminus \mathcal{X}_u$. The considered domain overlaid with the system's vector field is depicted in Fig. \ref{fig:vectorField}. Recall that our goal is to learn a control barrier function such that it's zero-sublevel set is a subset of $\mathcal{X}_\rho$.

By applying the proposed iterative, optimization-based method, we are able to obtain the control barrier function in the form of a collection of linear functions. For sampling, we took $M=200$ samples from the interior of $\mathcal{X}_\rho$. We set the input constraint to be $U_{feasible} = [-9,9]$. The optimization problem (\ref{Opt_Prob}) is solved using the CasADi solver (\cite{Andersson2019}) that uses an interior point method. The linear functions obtained are $h_1 = -0.75x_1 - 0.66x_2 -1.68$, $h_2 = 0.95x_1 - 0.32x_2 -2.36$, $h_3 =  0.981x_1 + 0.20x_2 -1.15$, $h_4 = -0.28x_1 + 0.96x_2 -1.26$, $h_5 = -0.07x_1 - 0.007x_2 -0.08$, and $h_6 = -0.20x_1 - 0.98x_2 -2.87$. The corresponding safe-invariant set is the intersection of half spaces of these linear functions, e.g.,
\begin{equation}
\mathcal{X}_{s}= \bigcap\limits_{i=1}^{6} \{ \boldsymbol{x} | h_i(\boldsymbol{x}) \geq 0 \}.
\end{equation}

\subsection{Safe Control using Learned CBF}
With the control barrier function obtained, we can now utilize the CBF-QP framework to implement a safe controller with constraints derived from the learned CBF. We first assume we have access to a nominal control input from a separate controller that does not guarantee system safety and can drive the system to dangerous states ($\mathcal{X}_u$ for instance). This controller can be a human controlling the system or an optimal controller that minimizes certain objectives. The goal is then to alter the nominal input in an minimally invasive way such that system safety is guaranteed. Below shows the CBF-QP framework

\begin{align}
    u(\boldsymbol x) &= \text{arg} \min \; \frac{1}{2} \lVert \Tilde{u}(x) - u \rVert ^2 \label{CBFQP1}\\
    &\quad s.t. \; \boldsymbol{A} \boldsymbol f(\boldsymbol x) + \boldsymbol{A} \boldsymbol g(\boldsymbol x)u \leq -\alpha(\boldsymbol{A}\boldsymbol x+\boldsymbol{b}) \label{CBFQP2}\\
    &\quad \quad \quad  \,   u_{min} \leq u \leq u_{max}
    \label{CBFQP3}
\end{align}

where $\boldsymbol{A}$ and $\boldsymbol{B}$ are as defined in (\ref{AandB}).

\begin{figure}
    \centering
    \includegraphics[width=0.4\textwidth]{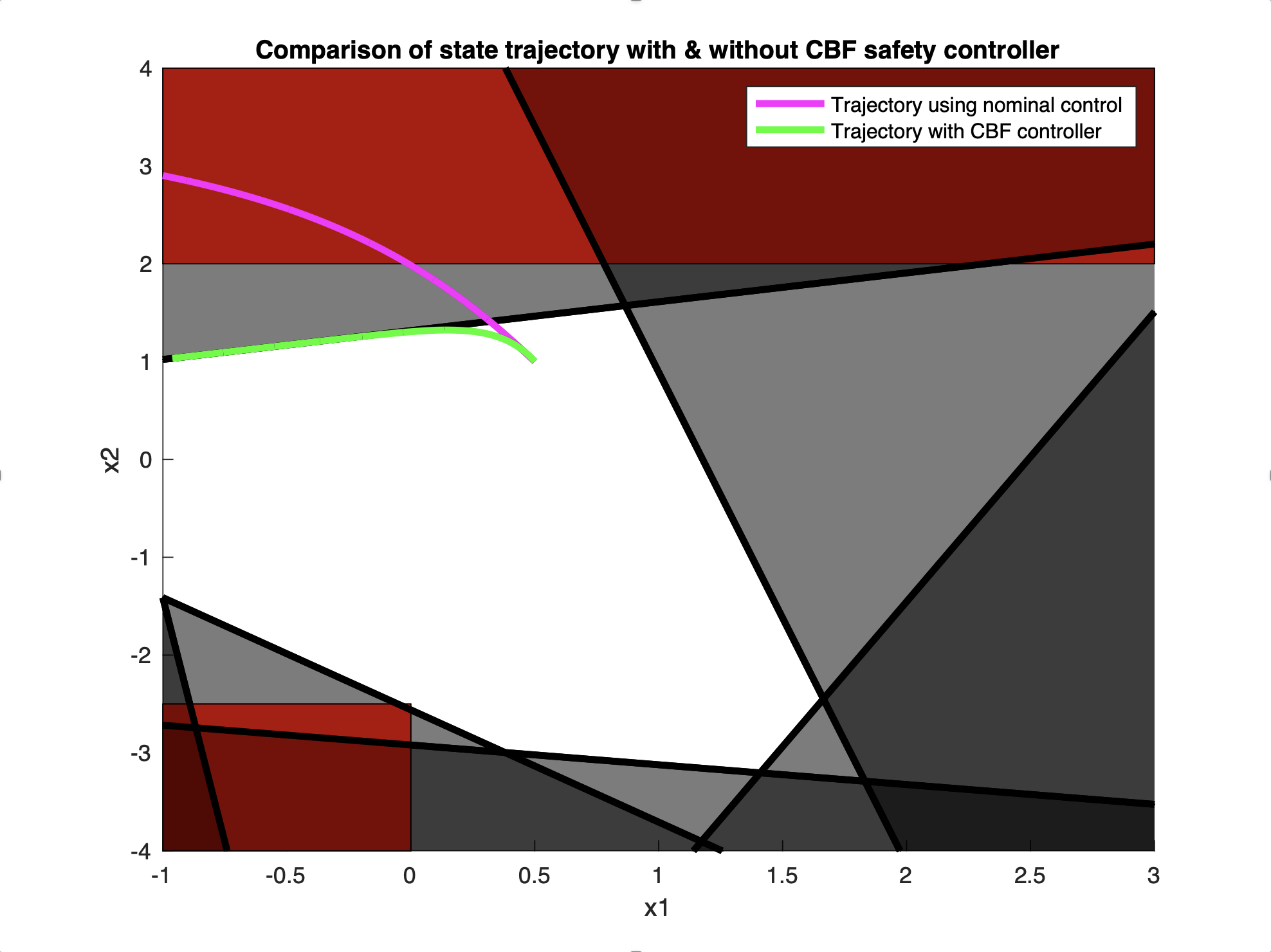}
    \caption{Comparison of state trajectory with and without CBF safe controller.}
    \label{fig:safeTraj1}
\end{figure}

\begin{figure}
    \centering
    \includegraphics[width=0.4\textwidth]{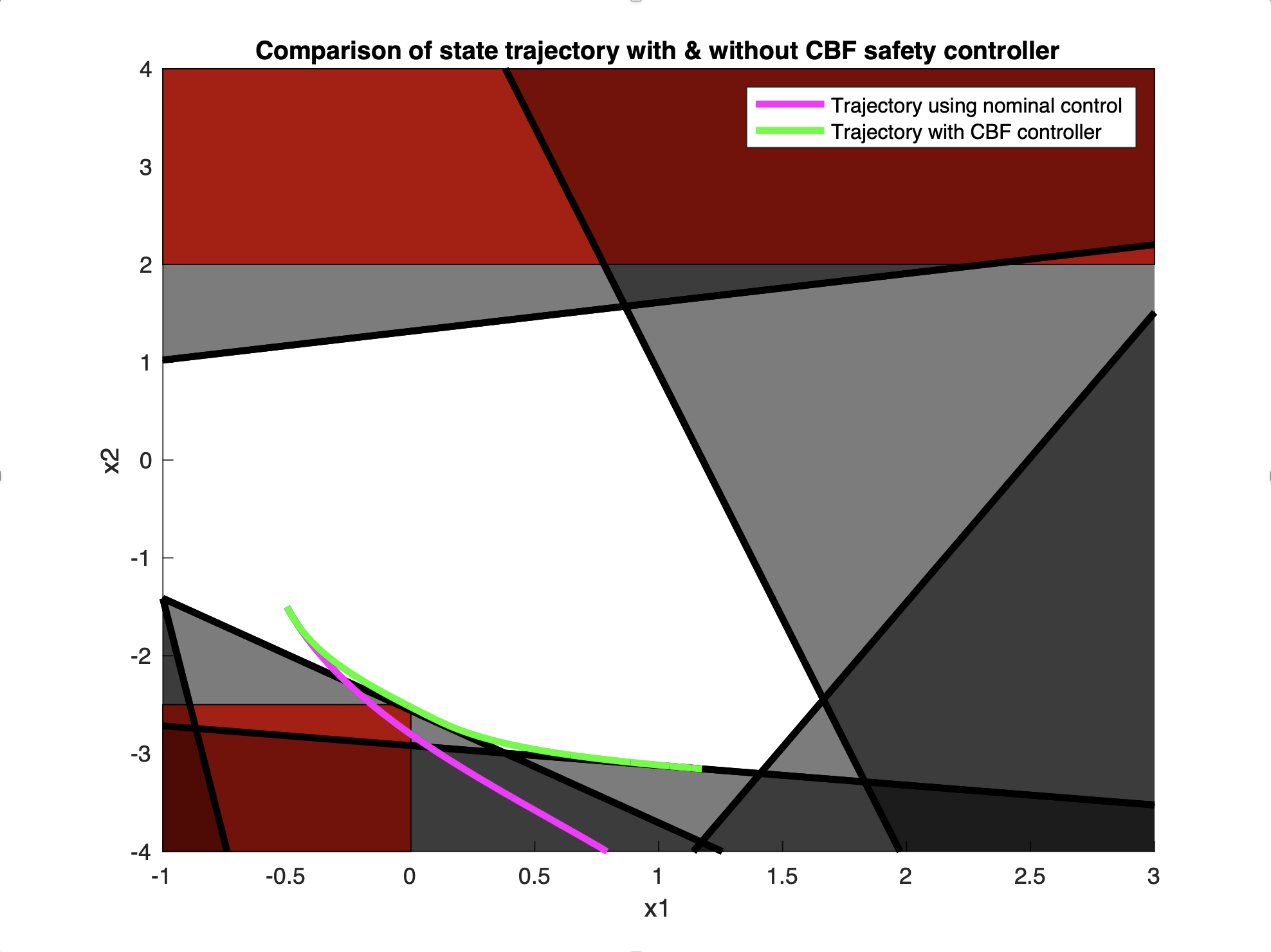}
    \caption{Comparison of state trajectory with and without CBF safe controller.}
    \label{fig:safeTraj2}
\end{figure}

The  obtained  safe  invariant  set  is  shown  in  Fig. \ref{fig:safeTraj1}.  The shaded region corresponds to the half-space $\{\boldsymbol{x}|h_i(\boldsymbol{x}) \leq 0\}$ and the unshaded region is the invariant set $\mathcal{X}_{s}$. We can see that $\mathcal{X}_{s}$ is a subset of $\mathcal{X}_\rho$ and that $\mathcal{X}_{s} \cup \mathcal{X}_u = \emptyset$. A nominal control $u_{nominal} = -3$ is applied to the system starting at $\boldsymbol{x}_0 = [0.5, 1]$. The CBF-QP framework (\ref{CBFQP1}-\ref{CBFQP3}) is then used to obtain a safe control. The magenta line and the green line in Fig. \ref{fig:safeTraj1} represents the system trajectories with and without the CBF-QP, respectively. We note that the nominal input drives the system into $\mathcal{X}_u$ since it does not concern about safety. On the other hand, the controller from CBF-QP kept the sytem within $\mathcal{X}_{s}$ via a smooth constrained trajectory, hence rendering the set forward invariant. Fig. \ref{fig:safeTraj2} denotes a similar case with system beginning at the initial state $\boldsymbol{x}_0 = [-0.5, -1.5]$.

\subsection{Effect of Input Constraints on Safe Set}
\begin{figure}
    \centering
    \includegraphics[width=0.35\textwidth]{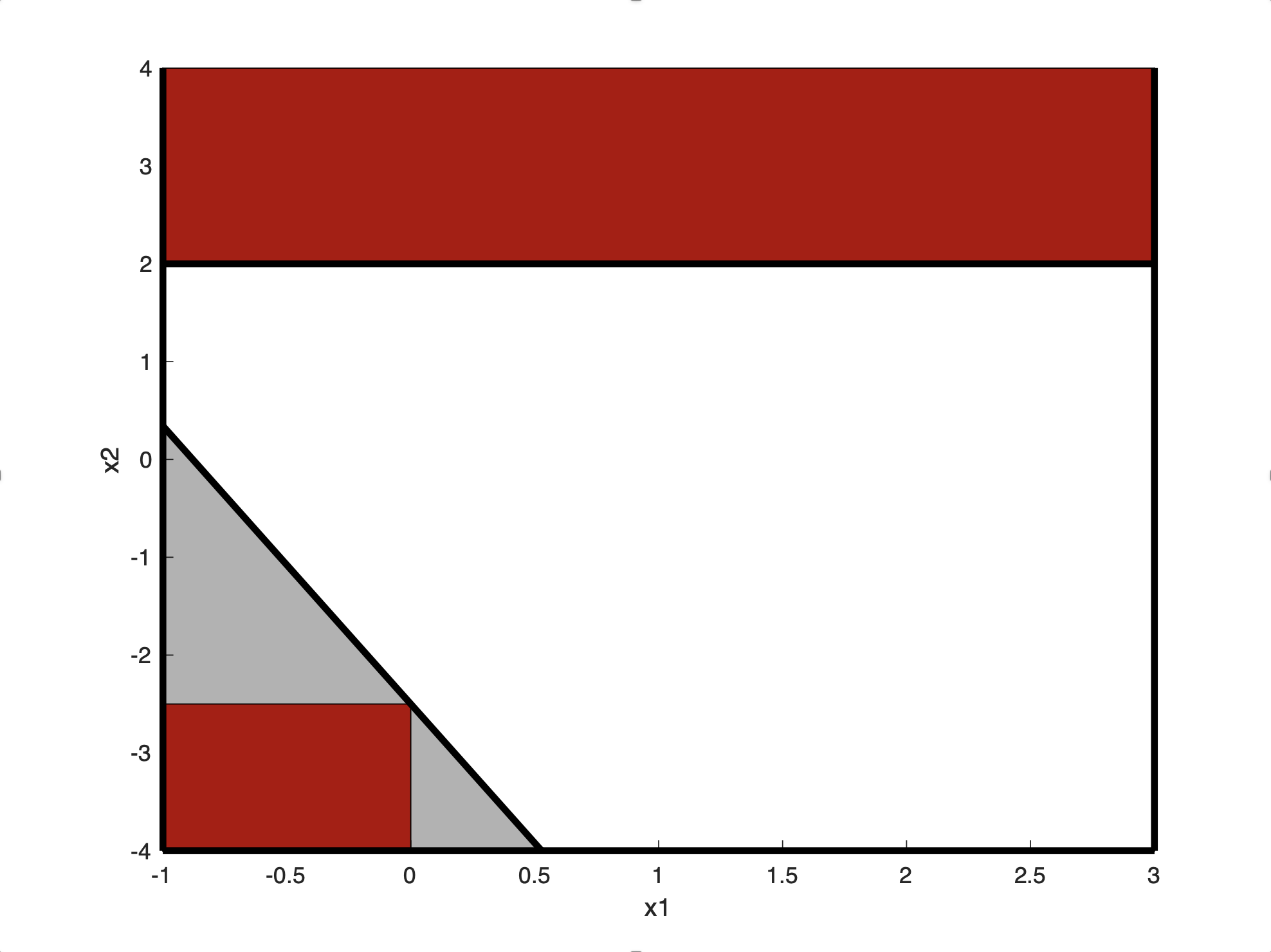}
    \caption{Safe set learned using $u_{min/max}=\pm\infty$}
    \label{fig:Large safe set}
\end{figure}

\begin{figure}
    \centering
    \includegraphics[width=0.35\textwidth]{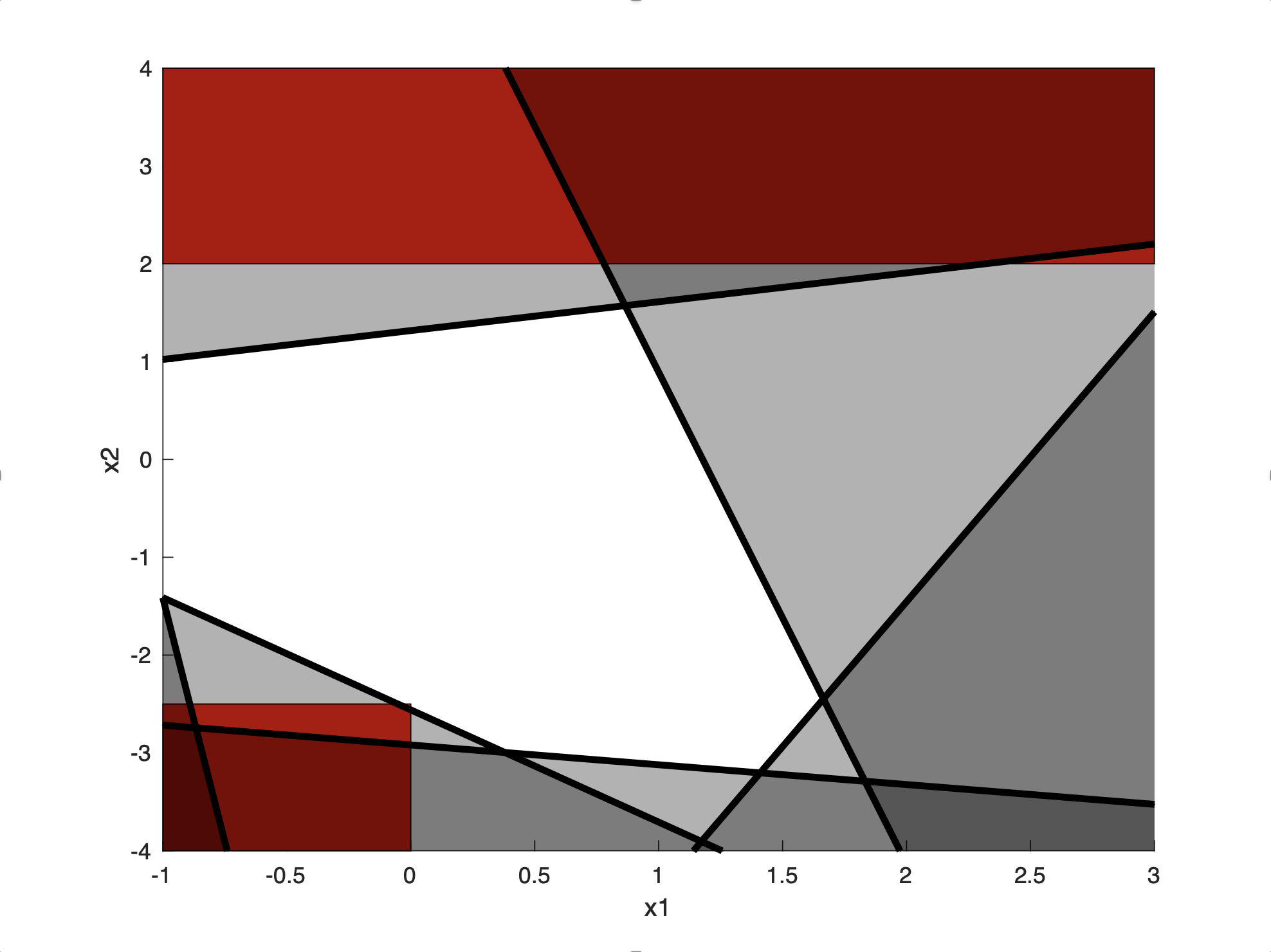}
    \caption{Safe set learned using $u_{min/max}=\pm 9$}
    \label{fig:Medium safe set}
\end{figure}

%Talk about how different input constraint range affects the volume of the learned safe set. 
Here, we elucidate how the bounds on control input of the system affects the control barrier function learned and the corresponding safe set. Fig. \ref{fig:Large safe set} and \ref{fig:Medium safe set} show different safe sets obtained with input bounds $u_{\text{min/max}}$ equals to $\pm \infty$, $\pm 9$, and $0$, respectively. In Fig. \ref{fig:Large safe set}, since the control is unbounded, this is equivalent to not enforcing constraint (\ref{CBFQP2}) in the optimization problem above. Recall that if the control is unbounded, we said the condition (\ref{h_dotcondition}) can always be satisfied (if the system is control-affine) by selecting a large magnitude input. Comparing this with Fig. \ref{fig:Medium safe set}, we can see the volume of the safe set is significantly smaller due to the control bounds. The shaded region in Fig. \ref{fig:Medium safe set} can be interpreted as states that cannot be render forward invariant by any input within the given bounds. Intuitively, these may be states that are too close to the unsafe states and the input required to drive the system away from $\mathcal{X}_u$ is well beyond the system's physical limit.

\section{Conclusion and Future Work}
In this work, we propose a iterative, optimization-based framework to obtain a control barrier function from a given set of allowable states defined by some application-specific safety measure. We show that given a nonlinear affine system with input constraints, the proposed method is able to obtain a set of linear functions and that the invariant set is indeed the intersection of half-spaces of these functions. We demonstrate the validity of the approach through a min-norm safe controller in which the constraints are derived using the obtained CBF. There are multiple directions that we would like to explore as future work. First, we wish to establish a formal proof that enable us to claim our approach is indeed maximizing the volume of the obtained safe set. Further, we would like to leverage the results in (\cite{ExpertDemonstrations}) and use Lipschitz properties to prove that, given the sampling is done at a fine enough resolution, if all states sampled satisfy our constraints, then even states that are not sampled will also satisfy such constraints. Another direction is to explore other representation such as sum-of-squares or ellipsoidal approximation in hope to improve the expressive power. Finally, we believe this work can be extended to a data-driven approach where we learn or correct the safe set in real-time as measurements are received.
\begin{comment}
    \begin{itemize}
        \item provable method to maximize volume 
        \item Explore other methods for parameterizing control barrier function
        \item Apply to other different applications and check performance
        \item Real-time learning/correcting of safe set with measurement-driven techniques
        \item Lipschitz sampling method
    \end{itemize}
\end{comment}

\bibliographystyle{unsrt}
\bibliography{ifacconf}             % bib file to produce the bibliography
                                                     % with bibtex (preferred)
                                                   
\end{document}